\documentclass[twocolumn,epsfig]{aa}
\usepackage{graphicx}
\begin{document}

\title{Parametric mechanism of the rotation energy pumping
by a relativistic plasma.}

\author{G.Z.Machabeli\inst{1}
   \and Zaza Osmanov\inst{2}\thanks{On leave from Abastumani
Astrophysical Observatory, Kazbegi ave.~2a, Tbilisi--380060,
Georgia}
   \and Swadesh M. Mahajan\inst{3}
  }
\institute{Abastumani Astrophysical Observatory, Kazbegi ave.~2a,
Tbilisi--380060, Georgia
 \and Dipartimento di Fisica Generale
dell'Universita degli studi di Torino, Via P. Giuria 1, I-10125,
Torino, Italy \and Institute for Fusion Studies, The university of
Texas at Austin, Austin, Texas 7871
} \offprints{Zaza Osmanov}

\date{Received / Accepted }
\authorrunning{Osmanov et al.}
\titlerunning{Parametric mechanism of the rotation energy pumping
by a relativistic plasma.}

\abstract{An investigation of the kinematics of a plasma stream
rotating in the pulsar magnetosphere is presented. On the basis of
an exact set of equations describing the behavior of the plasma
stream, the increment of the instability is obtained, and the
possible relevance of this approach for the understanding of the
pulsar rotation energy pumping mechanism is discussed.

\keywords{plasma, instabilities, pulsars, radiation} }

\maketitle

\section{introduction}

One of the important stages in the development of pulsar radiation
theory was the discovery that the rotation energy could transform
into energy in the electrostatic field; this possibility
stimulated the modeling of pulsar magnetospheres (see Ref.
~\cite{de},~\cite{gj}). In a typical model, the electrostatic
field $\bf E_0$, generated over the pulsar surface, has a finite
projection along the dipole magnetic field $\bf B_0$. The
resulting electric force, $e E_0$ exceeds the gravitational force
near the neutron star surface, and ends up uprooting charged
particles from the surface layer. These particles are accelerated
in the electric field and radiate $\gamma$ photons because of the
curvature radiation in the dipolar field. For photon energies
greater than twice the electron rest mass ( $\epsilon_\gamma\ge
2m_e c^2$),  the electron-positron pair production  ($\gamma+{\bf
B_0}\rightarrow e^+ +e^- +\gamma'$) in the ambience of the
magnetic field becomes possible. The particles, so produced,
repeat the cycle; they are accelerated, generate $\gamma$
radiation which creates more electron-positron pairs. This
cascading leads to the build-up of a large flux of relativistic
$e^+e^-$ pairs ; the process continues till the pair plasma
screens the electric field $\bf E_0$ as shown in Ref.
~\cite{st},~\cite{tad}.

The $e^+, e^-$ population in the pulsar atmosphere may be
conveniently divided into three principal components: 1) the basic
plasma mass (Bulk) with concentration $n_{pl}$, and a Lorentz
factor $\gamma_{pl}$, 2) a tail with concentration $n_t$, and
Lorentz factor $\gamma_t$, and 3) the remnant of the primary
electron or ion beam (most likely electrons). Note that  cascading
due to pair production is possible only in a highly relativistic
scenario, because for nonrelativistic or even mildly relativistic
velocities, the pair annihilation dominates pair production (see,
for example, Ref. ~\cite{dir},~\cite{kom}). The reader is referred
to Ref.~\cite{kaz},~\cite{mac} for a guide to processes that
define and control the behavior of radiation in the pulsar
magnetosphere.

The essence of this ``standard model'' is that the pulsar
magnetosphere is permeated with a multi-component plasma. It is
generally assumed that the energy contents of the three major
components are of the same order: $n_{pl}\gamma_{pl}\approx
n_t\gamma_t\approx n_b\gamma_b$. Theoretical estimates for the
density and the energy of the primary component are: $n_b\approx
(10^{11}-10^{12})\mbox{cm}^{-3}$ ( also called the Goldreich
Julian  concentration) and $\gamma_b\approx 10^6-10^7$. The origin
of the pulsar radiation is supposed to be in the magnetosphere,
and hence the energy contained in the $e^+e^-$ plasma must be
enough to account for the observable radiation.

In these pulsar models, the measure of the region over which the
electric field is nonzero is in great significance. The distance
between the star surface and the region where it is screened out
is called the vacuum gap, and may be estimated to be
$(10^3-10^4)\mbox{cm}$ for $E_0\approx 10^7\,$G(see for example
Ref. ~\cite{rud}). Unfortunately the  particle energy inventory
accumulated within the gap is not sufficient to explain the
observed visible radiation. Several mechanisms have been invoked
to increase the gap size. The intermediate formation of
positronium (electron-positron bound  state) which would hold back
the decay, and lead to an increase in the gap was proposed in
Ref.~\cite{us}. According to Ref.~\cite{aar}, the pulsar magnetic
field lines, supposedly curved towards rotation, must be screwed
even more; the twisted magnetic field lines will be rectified, and
as a result will enlarge the gap. A different mode, the PFF (pair
formation front) mechanism was introduced instead of the gap.
Taking the heating of the stellar surface and the electron
thermoemission into account leads to nonstationarity, and explains
nonstationary behavior of pulsars (see Ref.~\cite{kazm}).

In a series of papers, Muslimov and Tsigan have attempted to solve
the gap problem via the general relativity route. Realizing  that
in the vicinity of the rotating neutron star, the space-time is
slightly curved, they work out the creation of the electric field
in Kerr metric (see Ref.~\cite{mus}). As a result, the gap size is
somewhat enlarged but not enough to make a difference.

It seems that some additional source of energy or some new
mechanism will be necessary to surmount the problem arising out of
the ``insufficiency'' of the energy content in the vacuum gap. For
instance, it seems perfectly plausible and possible that one could
draw upon the pulsar rotation energy to augment the energy content
of the magnetospheric $e^+e^-$ plasma; the rotation energy, for
example,  could be transformed into   energy  associated with
oscillations in the $e^+e^-$ plasma far from the pulsar surface.
In this region the complicating effects of unremovable gravitation
(having to use, the Kerr metric, for example) will not exist. This
paper is an attempt to formulate and investigate the problem
associated with the parametric pumping of plasma oscillations in
the pulsar magnetosphere  taking into consideration the
complicated nature of the  multicomponent electron-positron
substrate.

Let us review a standard system to examine the plausibility of the
proposed mechanism. For the Crab Nebula, the nebula radiation can
be surely sustained by the rotational energy of the pulsar
PSR~0531 located in the vicinity of the nebular center. The power
of the nebula radiation exceeds the pulsar radiation power by two
orders of magnitude (it is approximately
$2\cdot10^{38}\,$erg/sec). The only source capable of providing
such a prodigious  power is the slowdown of the pulsar rotation:
$\dot W\approx I\Omega\dot \Omega$, where $I$ is pulsar's moment
of inertia with mass of order $(1.5-2.5)M_\odot$ ($M_\odot$ is
solar mass), $\Omega$ and $\dot\Omega$ are star's angular velocity
and angular acceleration respectively.  This power is equal to
$5\cdot 10^{38}\,$erg/sec.  The rotation energy loss rate is
described by the  ratio $\dot W/W\simeq2\dot\Omega/\Omega=2\dot
P/P$, where $P=2\pi/\Omega$ is the neutron star's rotation period,
and $P$ and $\dot{P}$ are measurable parameters for pulsars. Ratio
$\dot P/P$ for different pulsars ranges from $10^{-11}\sec^{-1}$
(PSR 0531) to $10^{-18}\sec^{-1}$ (PSR 1952+29). Thus, for
ordinary plasma oscillations to be pumped by the pulsar rotation,
their growth rate must not exceed $\dot P/P$, the measurable rate
of rotation energy loss.

A hydrodynamic approximation will be used to study the problem of
rotation induced wave generation in plasmas. For relative
simplicity one will assume that the  ${\it e}^+e^-$ plasma  has
two distinct energy ranges: the lower range (still highly
relativistic) with $n_{pl}$, and $\gamma_{pl}$, and the beam with
$n_b$, and $\gamma_b$. These flows propagate along the rotating
monopole like magnetic field lines.

\section{The main consideration}

It is supposed that for distances less than the radius of
curvature of the field line, the magnetic field is monopole like
(in this approximation the magnetic field lines may be supposed to
be rectilinear). The problem of the motion of charged particles in
the pulsar magnetosphere can be considered in the local inertial
frame of the observers, who measure the physical quantities in
their immediate vicinity. They are called the Zero Angular
Momentum Observers (ZAMOs). Naturally, the proper time of the
observer riding the particle, is different from the proper time of
ZAMOs.

Transformation from the inertial frame to the frame connected with
the pulsar magnetosphere can be done in the following way:
$$t = t', \varphi = \varphi',r = r',z = 0,\eqno(1)
$$ then the interval in the corotating frame will have the form:
$$
ds^2 = -\left(1-\Omega^2 r^2\right)dt^2-dr^2\eqno(2) 
$$ where $\Omega$ is angular velocity of rotation.

Note that the lapse function $\alpha = \sqrt{1-\Omega^2 r^2}$ ~($c
= 1 $) not only connects the proper time of ZAMOs with the
universal time $d\tau = \alpha dt$, but also gives a gravitational
potential:

$$ \bf g= -\frac{\bf\nabla\alpha}{\alpha}.\eqno(3) $$

The equation of motion of the particle in the ZAMOs frame is
expressed as:

$$\frac{d {\bf p}}{d\tau}=\gamma{\bf g}+\frac{e}{m}(\bf E+[\bf
V\bf
B])\eqno(4)
$$ where $\gamma = (1-\bf V^2)^{-1/2}$ is the Lorentz-factor and
${\bf V}=d{\bf r}/d\tau$ is the velocity of the particle
determined in the so-called 1+1 formalism, and ${\bf
p}\rightarrow{\bf p}/m$ is the dimensionless momentum.

As is shown in Ref.~\cite{chd} the transition from the particle
equation of motion to the Euler equation for fluid dynamics in the
$1+1$ formalism, may be easily fulfilled if one changes $d/d\tau$
in Eq. (4) by $1/(\alpha\partial t)+(\bf V\bf\nabla)$. The
resulting equation describing the stream motion (neglecting the
stream pressure) takes the following form: $$
\frac{1}{\alpha}\frac{\partial{\bf p}}{\partial t}+({\bf
V\nabla)p}=-\gamma\frac{\bf\nabla\alpha}{\alpha}
+\frac{e}{m}\left(\bf E+[\bf V\bf B]\right)\eqno(5) 
$$ where $\bf V$ and $\bf p$ are now hydrodynamic velocity and
momentum respectively. In order to rewrite this equation in the
inertial frame, let us note that the ZAMOs momentum coincides with
the momentum in the inertial frame. In fact, from the definitions
$\bf p= \gamma\bf V$, $\gamma=\alpha\gamma'$, and ${\bf V'}=d{\bf
r}/d\tau$ (prime refers to quantities in the inertial frame), one
can easily find that $\bf p= \bf p'$. In the inertial frame, then
Eq.~(5) converts to (omitting primes for all quantities):

$$ \frac{\partial{\bf p_i}}{\partial t}+({\bf v_i\nabla)p_i}=
-\gamma\alpha{\bf\nabla}\alpha+\frac{e_i}{m}\left(\bf E+
[\bf v_i\bf B]\right),\eqno(6) 
$$ $$i = b, e, p$$ where $b$, $e$ and $p$ denote the beam,
electron and positron components respectively. In Eq.(6), the
force $\bf F=-\gamma\alpha{\bf\nabla}\alpha$ is the analog of the
centrifugal force.

Adding the continuity and the Poisson equations:

$$ \frac{\partial n_i}{\partial t}+{\bf \nabla}(n_i{\bf
v_i})=0,\eqno(7)$$
$${\bf\nabla E}=4\pi e (n_e-n_p+n_b)\eqno(8) 
$$

As we are interested in the evolution of fluctuations, one can
look for solutions of Eqs.~({6})-({8}) in the framework of a
perturbation theory- an expansion in which terms like
$E^1/mn\gamma$ (the small parameter in the approximation of weak
turbulence for the plasma) are small: $$\bf E=\bf E^0+\bf
E^1+...\eqno(9a)$$ $$\bf B=\bf B^0+\bf B^1+...\eqno(9b)$$ $$\bf
p_i= p_i^0+ p_i^1+...\eqno(9c)$$ where $\bf E^0$, $\bf B^0$ and
$\bf p_i^0$ are the leading terms, and $\bf E^1$, $\bf B^1$ and
$\bf p_i^1$ constitute the perturbations. In the zeroth
approximation (taking into consideration the fact that the ejected
particles not only move along the radius, but also corotate with
the pulsar magnetosphere because of the {\it frozen-in} condition
, $\bf E_0+[\bf v_0\bf B_0]=0$) Eq.~({6}) will be reduced to the
form (see for example Ref.~\cite{mrog},~\cite{rig}):
\addtocounter{equation}{1}

$$
\frac{d^2r}{dt^2}=\frac{\Omega^2r}{1-\Omega^2r^2}\left(1-\Omega^2r^2-2
\left(\frac{dr}{dt}\right)^2\right)\eqno(10) 
$$ where we have neglected the term $(\bf v_0\nabla)\bf p_0$
(taking this term into account is a separate problem, and is not
examined within the framework of this paper). As is shown in
Ref.~\cite{mat}, Eq.~(10) allows an exact solution for particular
initial conditions $r(t_0 = 0) = 0$, $V(t_0 = 0) = V_0$:

$$ r(t)=\frac{V_0}{\Omega}\frac{Sn(\Omega t\mid \widetilde{m})}
{dn(\Omega t\mid \widetilde{m})}\eqno(11)
$$ where $Sn$ and $dn$ are Jacobian elliptical functions, the sine
and the modulus respectively and $\widetilde{m}=1-V_0^2$.

Using the following properties of the mentioned Jacobian
elliptical functions $Sn(x\mid 0) = sin(x)$ and $dn(x\mid 0) = 1$
(see Ref.~\cite{mat}), one may easily reduce Eq.~(11) for the
ultra-relativistic regime ($V_0\rightarrow 1$) relevant to this
paper:

$$
r(t)=\frac{V_0}{\Omega}\sin\Omega t.\eqno(12) 
$$

With this known asymptotic solution, the first order Eq.~(6)
reads:
$$\frac{\partial{\bf p_i^1}}{\partial t}+({\bf v_i^0\nabla)p_i^1=
F_i^1}+\frac{e_i}{m}\bf E^1,\eqno(13a)$$
$${\bf F_i^1}=\Omega^2 r v_i^0\bf p_i^1, \eqno(13b)$$
$$v_i^0=V_{i0}\cos\Omega t \eqno(13c)$$ where, in addition to
Eq.~(12),  we have used the Lorentz-factor expansion in the small
parameter $p_i^1/p_i^0$: $\gamma\approx \sqrt{1+\bf
p_i^{0^2}}\left(1+\bf p_i^0 p_i^1/(1+\bf p_i^{0^2})\right)$.

In a similar fashion, the linearized Eq.~(7) becomes:
\addtocounter{equation}{1}

$$ \frac{\partial n_i^1}{\partial t}+\mbox{div}(n_i^0{\bf
v_i^1})+\mbox{div}(n_i^1{\bf v_i^0})=0.\eqno(14)
$$

The electron-positron continuity equations may be combined
 to obtain the evolution  equation for the effective charge density
$n_{pl}=n_e-n_p$ :

$$ \frac{\partial n_{pl}^1}{\partial t}+\mbox{div}(n_e^0{\bf
v_{pl}^1})+\mbox{div}(n_{pl}^1{\bf
v_{pl}^0})=0\eqno(15) 
$$ where ${\bf v_{pl}^1}={\bf v_e^1}-{\bf v_p^1}$ and it has been
assumed that $\bf v_e^0=v_p^0\equiv v_{pl}^0$ and $n_e^0=n_p^0$.

We seek here a solution in which the density perturbations have no
spatial dependence so that the last terms of Eqs.~(14) and (15)
are identically zero. let us choose $n_i^1$ (now $i = pl, b$)  to
have the  form

$$
n_i^1=N_i e^{-\frac{ikV_{0i}}{\Omega}\sin\Omega t}.\eqno(16) 
$$

Then, if the rest of the  perturbed quantities are allowed the
spatial dependence $\bf p_i^1(\bf r;t)=\bf p_i^1(\bf k;t)e^{i\bf
k\bf r}$, one obtains from Eqs.~(14)-(15):

$$ ikp_i^1=-\frac{\gamma_{0i}^3}{n_i^0}\frac{\partial
N_i}{\partial
t}e^{-ikR_i}\eqno(17) 
$$ where $\gamma_{0i}$ is the initial Lorentz factor, and $R_i =
\frac{V_{0i}}{\Omega}\sin\Omega t$, and the relation
$v_i^1=p_i^1/\gamma_{0i}^3$ ($p_{pl}^1=p_e^1-p_p^1$) that is
satisfied for $\Omega t\sim\Omega/\omega \ll 1$ has been used.

Let us now go back to Eqs.~(13)  and write them separately for the
two components- the plasma and the beam : $$ \frac{\partial{\bf
p_b^1}}{\partial t}+({\bf v_b^0\nabla) p_b^1=F_b^1}+\frac{e}{m}\bf
E^1,\eqno(18a)$$ $$\frac{\partial{\bf p_{pl}^1}}{\partial t}+({\bf
v_{pl}^0\nabla)p_{pl}^1= F_{pl}^1}+2\frac{e}{m}\bf E^1.\eqno(18b)
$$

Repeating the  procedure applied to Eq.~(18) and taking Eqs.~(12)
and (17) into consideration, one can find that the beam and the
plasma components evolve as: $$\frac{\partial^2 N_b}{\partial t^2}
= -i\frac{en_b^0}{m\gamma_{b0}^3}e^{ikR_b}kE_1, \eqno(19a)$$
$$\frac{\partial^2 N_{pl}}{\partial t^2} =
-2i\frac{en_{pl}^0}{m\gamma_{p0}^3}e^{ikR_{pl}}kE_1. \eqno(19b)$$

Combining Eqs.~(19a) and (19b) eliminates the electric field to
yield \addtocounter{equation}{2}

$$ \frac{\partial^2 N_b}{\partial t^2}=
\frac{n_b^0\gamma_{p0}^3}{2n_{pl}^0\gamma_{b0}^3}e^{ik(R_{pl}-R_b)}
\frac{\partial^2 N_{pl}}{\partial t^2}.\eqno(20) 
$$ Equation (20) is a rather complicated non-autonomous equation
in time. To solve it, one can take its Fourier  time transform
(restoring the speed of light) along with that of the Poisson
Eq.(8) [cf. Appendixes~A,B Eqs.~(A.2) and (B.2)] with the electric
field eliminated to arrive at the coupled system

$$ \omega^2N_b(\omega)=\frac{n_b^0\gamma_{0pl}^3}
{2n_{pl}^0\gamma_{0b}^3}\sum_s(\omega+s\Omega)^2 J_s(a)
N_{pl}(\omega+s\Omega),\eqno(21) $$ $$
\left(\omega^2-\frac{\omega_{pl}^2}{\gamma_{0pl}^3}\right)
N_p(\omega) = \frac{\omega_{pl}^2}{\gamma_{0pl}^3}\sum_s
J_s(a)N_b(\omega-s\Omega)\eqno(22) $$ where $a
=kc/2\Omega\gamma^2_{0pl}$, and $\omega_{pl} = \sqrt{8\pi
n_{pl}^0e^2/m}$. Naturally one can see the appearance of
convolution sums on the right hand sides of  both equations.

Substituting $N_b$ from Eqs.~(21) into (22), one can find:

$$ \left(\omega^2-\frac{\omega_{pl}^2}{\gamma_{0p}^3}\right)
N_{pl}(\Omega)=$$ $$=\frac{\omega_b^2}{\gamma_{0b}^3}
\sum_{sl}J_s(a)J_l(a)\left(\frac{\omega-(s-l)\Omega}
{\omega-s\Omega}\right)^2N_{pl}(\omega-(s-l)\Omega)\eqno(23) 
$$ where $\omega_b = \sqrt{8\pi n_b^0e^2/m}$. One could try to
find the nature of the time evolution from Eq.(23), but it seems
to be a little better to go back to $n_{pl}$ given by
$N_{pl}=n_{pl}^1e^{ikR_{pl}}$.  Carrying out the algebra given in
the appendix[cf. Appendix~C, Eq.~(C.2)],  one may derive

$$
N_{pl}=\sum_{s=-\infty}^{+\infty}J_s(b)n_{pl}(\omega-s\Omega)\eqno(24)
$$ where $b = kc/\Omega$. Substituting Eq.(24 ) into Eq.~(23), one
finally obtains the rather complicated dispersion: $$
\left(\omega^2-\frac{\omega_{pl}^2}{\gamma_{0p}^3} \right)\sum_s
J_s(b)n_{pl}(\omega-s\Omega)=$$ $$\frac{\omega_b^2}{\gamma_{0b}^3}
\sum_{jlm}J_l(a)J_{j+l+m}(a)J_m(b)\left(\frac{\omega+(j+m)\Omega}
{\omega-l\Omega}\right)^2n_{pl}(\omega+j\Omega).\eqno(25) $$

To extract  some sense  out of the above result, one can explore
the the dispersion relation near the resonant condition,
$\omega^2\approx\omega_{pl}^2/\gamma_{0pl}^3$. Near the resonance,
the basic contribution to the sum
$\sum_{s}J_s(b)n_{pl}(\omega-s\Omega)$ comes from $\omega\approx
s_0\Omega$. Similarly the right hand side of (25) is reduced to a
single term corresponding to $\omega\approx -j_0\Omega$,
$\omega\approx l_0\Omega$ and $m = m_0=-j_0=l_0=s_0$. Rewriting
$\omega^2-\omega_{pl}^2/\gamma_{0pl}^3$  as
$2\omega_{pl}\Delta/\gamma_{0pl}^{3/2}$ (where
$\Delta=\omega-\omega_{pl}/\gamma_{0pl}^{3/2}$), then, reduces
Eq.~(25) the simple cubic equation:

$$\Delta^3\approx\frac{\omega_b^2\omega_{pl}}{2
\gamma_{0b}^3\gamma_{0pl}^{3/2}}J_{s_0}^2(a) \eqno(26) $$ where
$s_0=[\omega_{pl}/\Omega\gamma_{0pl}^{3/2}]$. In addition to the
real root, the dispersion relation of Eq.~(26) allows the complex
conjugate pair

$$ \Delta_{1,2}\approx-\frac{1}{2}M\pm\frac{i\sqrt
3}{2}M,\eqno(27a) $$

$$ M=\left[\frac{\omega_b^2\omega_{pl}}{2
\gamma_{0b}^3\gamma_{0pl}^{3/2}}J_{s_0}^2(a)\right]^{1/3}\eqno(27b)
$$ with comparable real and imaginary parts. The root with the
positive imaginary part implies the instability that we were
seeking.
    \begin{figure}[hbt]
 \par\noindent
 {\begin{minipage}[t]{1.\linewidth}
 \includegraphics[width=\textwidth] {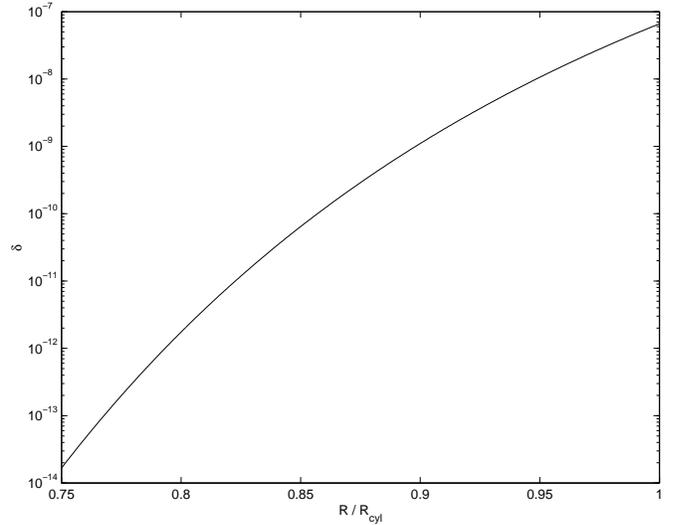}
 \end{minipage}
 }
 \caption[] {Graph of $log(\delta(R / R_{cyl}))$. Set of parameters is
following: $n_{0b}\approx 10^7\mbox{cm}^{-3}$, $\gamma_{0b}\approx
10^6$, $\gamma_{0pl}\approx 10$, $\Omega\approx 190\,$Hz,
$r\approx 10^6cm$.}\label{FIG}
 \end{figure}

\section{estimates}

The instability turns out to be rather strong. For typical pulsar
parameters (that of the Crab nebula) $n_{0b}\approx
10^7\mbox{cm}^{-3}$, $\gamma_{0b}\approx 10^6$,
$\gamma_{0pl}\approx 10$, $\Omega\approx 190$Hz, considering the
longitudinal waves, one can plot the graph of the growth rate
$\delta$ ($\delta = \sqrt{3}M/2$ see Eq.~(27)) as a function of
the radial distance (normalized by the light cylinder radius). For
simplicity let us examine longitudinal waves in a coupling point.
In this case all three modes of a cold plasma (o, x and Alfven
modes) are indistinguishable, then one can estimate the wave
number according to the following approximate formula
$kc\approx\omega_{pl}/\gamma_{0pl}^{3/2}$ (see Ref.~\cite{mmel}).
Using mentioned pulsar parameters, from Eq. (27a) the growth rate
is estimated. In Fig.1 one may see that $\delta$ is sensitive to
the radial distance, in the range $0.75\leq R/R_{cyl} \leq 1$ it
increases from $\sim10^{-14}$ to $\sim10^{-7}$. (here $R_{\rm
cyl}$ is the light cylinder radius, thus the radius, where the
rotation velocity equals to the speed of light). It is clear that,
$\delta$ becomes unreasonably large for $R/R_{cyl}\geq 0.85$,
which means that, the linear assumption will be grossly violated
long before these distances are reached. The instability will have
a linear growth over certain distances and then nonlinear
saturation will set in after which other nonlinear mechanisms may
take place in the energy pumping phenomenon. In estimating the
growth one has invoked $n_{0pl}\gamma_{0pl}\approx
n_{0b}\gamma_{0b}$, and used the fact that the density goes as
$n\approx n_0 (r/R)^3$ (where $r\approx 10^6 cm$ is a radius of
the neutron star) because the magnetic filed is monopole like.

\section{conclusions}

The purpose of this paper was to explore  the possibility of
pumping the  rotational energy of the pulsar into the plasma.
Considering a highly idealized system, the linear instability
caused by rotation in a two component relativistic plasma embedded
in a uniform magnetic field was examined. By using the
hydrodynamic equations of motion, the continuity and the Poisson
equation, it has been shown that the plasma waves can grow on the
rotation energy with rather high growth rates. In fact the
perturbation growth rate for distances $R/R_{cyl}\geq 0.85$ is
quite high in comparison with the supposed  instability rates we
were seeking. If this mechanism is, indeed, operational, then  the
only consistent scenario is that the linear stage with this growth
rate is very short, and nonlinearities are turned in soon enough
to considerably reduce the growth. Thus the need for a nonlinear
theory is immediately and strongly indicated. Sooner or later we
should consider this particular nonlinear effect, which will
comprise one more step closer to the real scenario. There are two
other shortcomings of this effort: a) the straight magnetic field
lines has been examined, whereas real profiles are curved and b)
only electrostatic waves have been considered. The preliminary
results, however, unambiguously show that the energy content of
the magneto spheric plasma can grow at the expense of the stellar
rotational energy.

\section{ acknowledgements}
The study of SMM  was supported by US Department of Energy
Contract No.DE-FG03-96ER-54366.

\appendix

\section{Derivation of Eq.~(21)}

Note that: $$N_f(t)=\int d\omega' e^{-i\omega't}N_f(\omega')
\eqno(A.1a)$$ where $$f=b,pl$$ then $$\int dt
e^{i\omega't}\frac{\partial^2}{\partial t^2}\int d\omega'
e^{-i\omega't}N_f(\omega')=$$ $$=\int\int dtd\omega'e^{i\omega
t}(-i\omega')^2e^{-i\omega't}N_f(\omega')=$$ $$=-\int\int
dtd\omega'e^{it(\omega-\omega')}(\omega')^2N_f(\omega')=$$
$$=-2\pi\int d\omega'
(\omega')^2\delta(\omega-\omega')N_f(\omega')=$$
$$=-2\pi\omega^2N_f(\omega) \eqno(A.1b)$$ where the following
representation of the delta function has been examined:
$$\delta(x)=\frac{1}{2\pi}\int dk e^{ikx}, \eqno(A.1c)$$
$$R_{pl}-R_b=\frac{V_{0pl}-V_{0b}}{\Omega}\sin\Omega t\approx$$
$$\approx\frac{c}{\Omega}\left(\frac{1}{2\gamma_{0b}^2}-\frac{1}{2\gamma_{0pl}^2}
\right)\sin\Omega
t\approx-\frac{c}{2\gamma_{0pl}^2\Omega}\sin\Omega t,
\eqno(A.1d)$$ then $$e^{-ik(R_{pl}-R_b)}\approx e^{ia \sin\Omega
t},\eqno(A.1e)$$
$$a=\frac{kc}{2\gamma_{0pl}^2\Omega}.\eqno(A.1f)$$

We have considered the approximate expression: $V_{0i}\approx
1-1/2\gamma_{0i}^2$, and the following observable fact:
$\gamma_{0b}\gg\gamma_{0pl}$. As one can see, the speed of light
has been restored again.

By using the following identity: $$e^{\pm ix\sin\Omega t}=\sum_s
J_s(x)e^{\pm is\Omega t},\eqno(A.1g)$$ one may easily transform
$e^{-ik(R_p-R_b)}$ into its Fourier mode: $$\int dt e^{i\omega
t}\sum_s J_s(a)e^{is\Omega t} \frac{\partial^2}{\partial t^2}\int
d\omega' e^{-i\omega' t}N_p(\omega')=$$ $$=-\sum_s J_s(a)\int\int
dtd\omega'(\omega')^2N_p(\omega')e^{it(\omega+s\Omega-\omega')}=$$
$$=-2\pi\sum_s J_s(a)\int
d\omega'(\omega')^2N_p(\omega')\delta(\omega+s\Omega-\omega')=$$
$$=-2\pi\sum_s
J_s(a)(\omega+s\Omega)^2N_p(\omega+s\Omega).\eqno(A.1h)$$

Combining Eqs.~(A.1b) and (A.1h), taking into consideration
Eq.~(20), one finally will obtain: $$\omega^2
N_b(\omega)=\frac{n_b^0\gamma_{0pl}^3}
{2n_{pl}^0\gamma_{0b}^3}\sum_{s}(\omega+s\Omega)^2
J_s(a)N_{pl}(\omega+s\Omega).\eqno(A.2)$$

\section{Derivation of Eq.~(22)}

One can easily transform Eq.~(8) into the following form: $${\bf k
E_1}=4\pi e\left(N_b(t)e^{-ikR_b}+N_{pl}(t)
e^{-ikR_{pl}}\right).\eqno(B.1a)$$

Taking Eq.~(19b) into account, one can have: $$\frac{\partial^2
N_{pl}(t)}{\partial t^2}=-\frac{8\pi en_{pl}^0}{m\gamma_{0pl}^3}
\left(N_{pl}(t)+N_b(t) e^{ik(R_{pl}-R_b)}\right).\eqno(B.1b)$$

For Fourier expansion of the left-hand side of Eq.~(B.1b) one can
analogously go to Eq.~(A.1b) write down: $$\int dt
e^{i\omega't}\frac{\partial^2}{\partial t^2}\int d\omega'
e^{-i\omega't}N_{pl}(\omega')=-2\pi\omega^2N_{pl}(\omega)
\eqno(B.1c)$$ while, for the second term in a bracket, one finds:
$$\sum_s J_s(a)\int\int
dtd\omega'e^{it(\omega-s\Omega-\omega')}N_b(\omega')=$$
$$=-2\pi\sum_s J_s(a)\int
d\omega'N_b(\omega')\delta(\omega-s\Omega-\omega')=$$
$$=-2\pi\sum_s J_s(a)N_b(\omega-s\Omega).\eqno(B.1d)$$

Combining Eqs.~(B.1c), and (B.1d) we will obtain Eq.~(22):
$$\left(\omega^2-\frac{\omega_{pl}^2}{\gamma_{0pl}^3 }\right)
N_{pl}(\omega) = \frac{\omega_{pl}^2}{\gamma_{0pl}^3 }\sum_{s}
J_s(a)N_b(\omega-s\Omega) \eqno(B.2a)$$ where $$\omega_{pl} =
\sqrt{\frac{8\pi n_{pl}^0e^2}{m}}.\eqno(B.2b)$$

\section{Derivation of Eq.~(24)}

From Eq.~(16) we have:

$$N_f(t)=n_f^1(t)e^{ikR_f}.\eqno(C.1a)$$

Taking into consideration Eq.~(A.1g), and a condition $\gamma_a\gg
1$ one finds: $$N_f(t)=\sum_s J_s(A_f)e^{is\Omega t}n_f^1(t)
\eqno(C.1b)$$ where $$A_b=\frac{kc}{\Omega}\eqno(C.1c)$$
$$A_{pl}=\frac{kc}{2\Omega \gamma_{0pl}^2}.\eqno(C.1d)$$

Fourier analysis of the right-hand side of the Eq.~(C.1b) gives
following: $$\sum_s J_s(A_f)\int\int
dtd\omega'e^{it(\omega+s\Omega-\omega')}n_f(\omega')=$$
$$=-2\pi\sum_s J_s(A_f)\int
d\omega'n_f(\omega')\delta(\omega+s\Omega-\omega')=$$
$$=-2\pi\sum_s J_s(A_f)n_f^1(\omega-s\Omega)\eqno(C.1e)$$ Hence we
will have: $$N_f(\omega) =
\sum_sJ_s(A_f)n_f^1(\omega-s\Omega)\eqno(C.2)$$

\end{document}